\documentclass[preprint,showpacs,showkeys,amsmath,amssymb]{revtex4}

\usepackage{graphicx}
\usepackage{dcolumn}
\usepackage{bm}

\begin{document}


\title{Gravity model in the Korean highway}

\author{Woo-Sung Jung}
\email{wsjung@physics.bu.edu}
\affiliation{Center for Polymer Studies and Department of Physics, Boston University, Boston, MA 02215, USA}

\author{Fengzhong Wang}
\affiliation{Center for Polymer Studies and Department of Physics, Boston University, Boston, MA 02215, USA}
\author{H. Eugene Stanley}
\affiliation{Center for Polymer Studies and Department of Physics, Boston University, Boston, MA 02215, USA}

\date{\today}

\begin{abstract}
We investigate the traffic flows of the Korean highway system, which contains both \textit{public} and \textit{private} transportation information. We find that the traffic flow $T_{ij}$ between city $i$ and $j$ forms a gravity model, the metaphor of physical gravity as described in Newton's law of gravity, $P_iP_j/r_{ij}^2$, where $P_i$ represents the population of city $i$ and $r_{ij}$ the distance between cities $i$ and $j$. It is also shown that the highway network has a heavy tail even though the road network is a rather uniform and homogeneous one. Compared to the highway network, air and public ground transportation establish inhomogeneous systems and have power-law behaviors.
\end{abstract}

\pacs{89.75.Hc, 89.40.Bb, 89.75.-k}

\keywords{Highway, Traffic flows, Gravity model}

\maketitle

\section{Introduction}
Complex network science has been an active interdisciplinary research field and the methodology of physics is applied into many interdisciplinary researches \cite{mantegna1999B,arthur1997B,bouchaud2000B,weidlich2000B,watts1999B,barabasi2002B,jkps}. Numerous real-world networks have been investigated, and scaling laws and patterns have been observed in nature, society, and so on \cite{watts1999,barabasi99,newman03}. Transportation network has attracted a lot of interests, particularly many public transportation systems. Amaral \textit{et al.}, Barrat \textit{et al.}, and Guimer\`a \textit{et al.} \cite{amaral2000,barrat2004,guimera2005} have found small-world behaviors and truncated power-law cumulative degree distribution $P(k)\propto k^{-\alpha}f(k/k_x)$ with the exponent $\alpha=1.0$ from world-wide airport network. Moreover, the airport networks of India \cite{bagler2004} and China \cite{li2004} have shown small world behaviors. The Indian and Chinese airport networks are characterized by small average path lengths ($\left< l\right> \approx 2$) and large clustering coefficients ($c>0.6$). However, the degree distribution of the Indian network follows a power-law and that of the Chinese a truncated power law. Latora \textit{et al.} \cite{latora2002} studied the Boston public transportation system including subway and bus, and Seaton \textit{et al.} \cite{seaton2004} compared the underground system of Boston ti that of Vienna. The network efficiency defined as a mean value of inverse distances between nodes has been analyzed, and that of the two systems, Boston and Vienna, shows a good correspondence regarding the value of average degree, but other properties such as clustering coefficient and network size have shown differences. Sienkiewcz \textit{et al.} \cite{sienkiewicz2005} analyzed the public transport systems in Poland and observed similar features with other works. Brockmann \textit{et al.} \cite{brockman2006} studied human traveling statistics in the United States by analyzing the circulation of bank notes. They described the dispersal bank notes and human travels by a continuous-time random-walk process that incorporates scale-free jumps as well as long waiting times between displacements.

At present, the studies on the traffic systems have usually been focused on the public transportation system. Collecting the data for the public system is much easier than that for private travels. In fact, it is almost impossible to keep tracking of each traveler. However, the Korean highway has a special toll system. First of all, all Korean highway are toll roads. Each and every exit has a toll plaza, and every car's traveling information including the exit number that comes in and goes out is reported. Therefore, we can keep tracking of the travel information of a single car. 

We investigate the traffic flows on the Korean highway, and identify a gravity model. The traffic flow $T_{ij}$ between city $i$ and $j$ is proportional to $P_iP_j/r_{ij}^2$, where $P_i$ represents the population of city $i$ and $r_{ij}$ the distance between two cities, $i$ and $j$. We also find that the highway network including both the information of the public and private transportation has a heavy tail but not the power-law behavior, even though the road network is a rather uniform and homogeneous one like random network \cite{jeong2003}, compared to air and public ground transportation networks which establish inhomogeneous systems and have power-law behaviors \cite{barrat2004,montis2005}.

\section{The Korean highway system}
We investigate the traffic flows between cities on the Korean highway system for the year of 2005 \cite{koreaex}. The system consists of 24 routes and 238 exits. The total length of the system is 3,050km, and the longest route, connecting Seoul and Busan, is about 418km \cite{moct}. The resulting highway network comprises 238 nodes denoting exits and fully connected links to each other. The \textit{Republic of Korea}, usually called \textit{South Korea}, is not a big country, whose area is 99,646km$^2$ \cite{moct,koreaindex}. It occupies only the southern portion of the Korean Peninsula, and borders on only North Korea. The two Koreas do not have any public road connecting each other. Therefore, South Korea is a sort of island from the viewpoint of road system. This property limits our statistical analysis in comparison with larger country or continent such as the United States or the European Continent.

First of all, we select 30 cities, whose population is over 200,000, for analysis (Section III). They are as shown in Table \ref{table:city}. Other studies exploring the transportation networks usually deal with small number of nodes, such as N=79 \cite{bagler2004}, 128 \cite{li2004}, N=124 \cite{latora2002}, and N=76 \cite{seaton2004} . Only a few surveys such as the Indian Railway Network (N=579) \cite{sen2003}, and the Worldwide Airport Network (N=3880) \cite{barrat2004} deal with larger number of nodes. A small number of cities, 30, may restrain statistical analysis. However, it is noticed that this is an empirical analysis for such private traffic flows, which is difficult to be collected. 

\begin{table*}
\begin{center}
\begin{tabular}{|c|c|c|c||c|c|c|c|}
\hline
\#&City&Population&Area (km$^2$)&\#&City&Population&Area (km$^2$)\\
\hline
1&Seoul&9,762,550&605.52&16&Masan&426,784&329.69\\
2&Busan&3,512,550&763.36&17&Gumi&381,583&616.24\\
3&Daegu&2,456,020&885.70&18&Pyeongtaek&374,262&452.13\\
4&Daejeon&1,438,550&539.66&19&Jinju&336,355&712.62\\
5&Gwangju&1,413,640&501.35&20&Iksan&306,974&507.11\\
6&Ulsan&1,044,930&1,056.74&21&Wonju&283,583&867.22\\
7&Suwon&1,039,230&121.05&22&Gunpo&268,917&36.39\\
8&Yongin&686,842&591.47&23&Gyeongju&266,131&1,324.00\\
9&Ansan&669,839&147.14&24&Suncheon&261,519&907.43\\
10&Cheongju&640,631&153.41&25&Chuncheon&260,234&1,116.27\\
11&Jeonju&622,092&206.24&26&Gunsan&249,212&380.06\\
12&Cheonan&518,171&636.29&27&Mokpo&244,543&47.92\\
13&Changwon&499,414&292.65&28&Gyeongsan&240,371&411.57\\
14&Pohang&488,433&1,127.74&29&Gangneung&220,706&1,040.2\\
15&Gimhae&428,893&463.33&30&Chungju&204,248&984.07\\
\hline
\end{tabular}
\caption{\label{table:city} Thirty largest cities in Korea for the year of 2005. They are shown in an alphabetic order. Population is the actual population from the official Korean government census \cite{koreaindex}. This \textit{door-to-door} census may have less data than real population. Area is also measured by the government \cite{moct,koreaindex}.}
\end{center}
\end{table*}

Second of all, we analyze the whole traffic data with a total of 238 exits (Section IV). The data of the whole exits can provide more statistical analysis, but not directly related to population or area. However, the data analyzed do not include either the actual path of transportation, or the number of transportations which go along the path connecting with the two nearest exits. 

\section{Gravity model}
\begin{figure*}
\begin{center}
\includegraphics[width=0.8\textwidth]{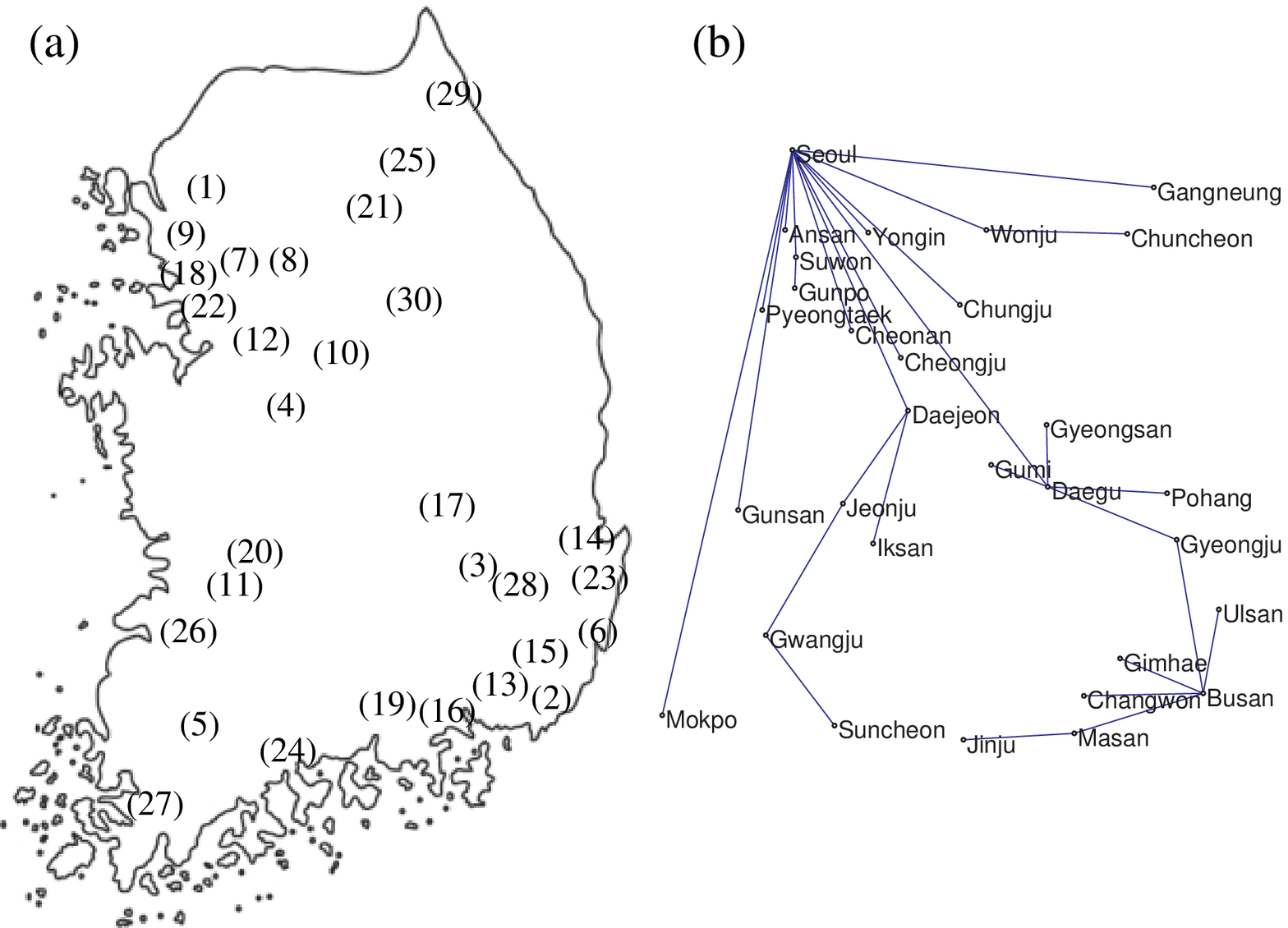}
\caption{(a) Map of Korea and thirty selected cities. (b) The maximum spanning tree of traffic flows consisting of 30 cities.}
\label{fig:map}
\end{center}
\end{figure*}

In this section, we study the traffic flows and gravity model from the data of 30 selected cities. Fig. \ref{fig:map}(a) illustrates 30 cities on the map. Each number from 1 to 30 corresponds to the city number of Table \ref{table:city}. A minimum spanning tree (MST) is a useful tool to analyze complex network \cite{mantegna1999,kullmann2000,jung2006}. When a weight is assigned to each link of network, an MST is constructed with a weight less than the weight of every other spanning tree. The highway network, which a node represents a \textit{city}, is established. Some cities have many exits. We consider the whole exits of a city as one exit for the city in this section. This network is a weighted one, wherein the weight $w_{ij}$ of a link connecting city $i$ and $j$ represents the sum of both traffic flows $i\to j$ and $j\to i$. In the highway system, the important link is the one that has larger traffic flow than others. Therefore, we construct the \textit{maximum} spanning tree of the highway system (Fig. \ref{fig:map}(b)).

\begin{figure}
\begin{center}
\includegraphics[width=0.8\textwidth]{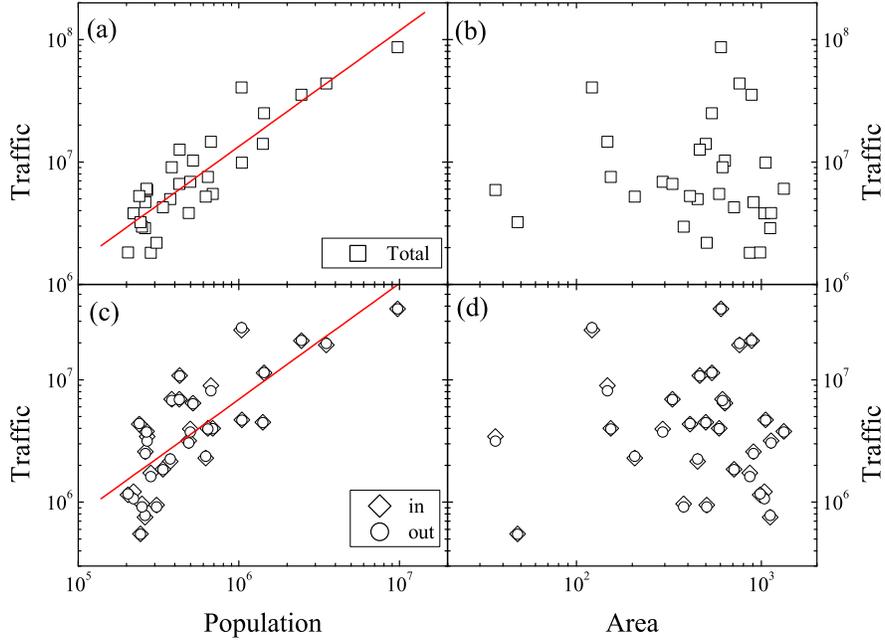}
\caption{Correlation between total traffic and (a) the population, (b) the area. The traffic is proportional to the population, and the correlation coefficient is 0.95 on a log-log scale. However, the area is not correlated with the traffic. (c) and (d) represent correlations between inbound, out bound traffic flows, and the population, the area. Both flows are proportional to the population, and the correlation coefficient is 0.93 (inbound) and 0.95 (outbound), respectively.}
\label{fig:population}
\end{center}
\end{figure}

\begin{figure}
\begin{center}
\includegraphics[width=0.8\textwidth]{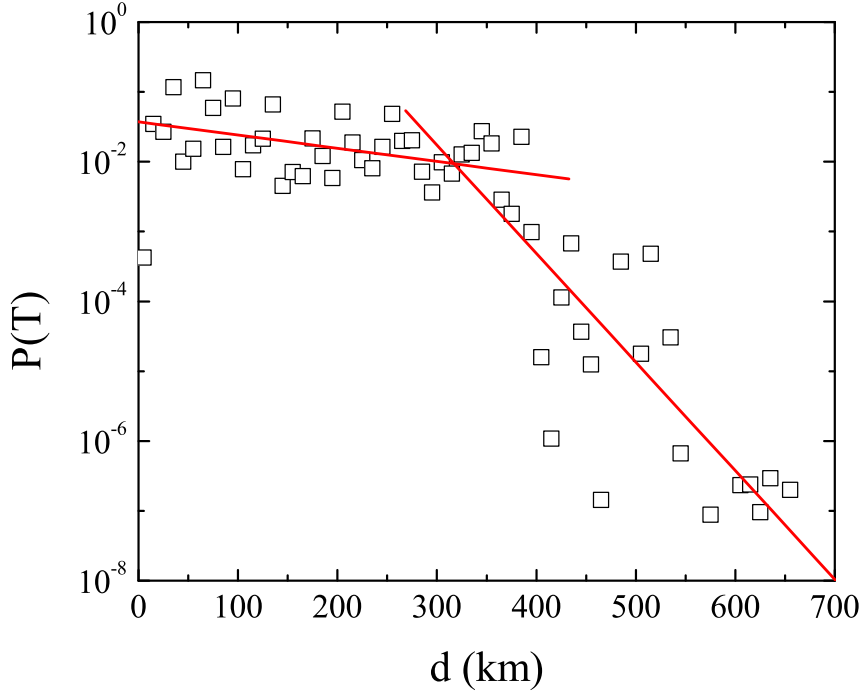}
\caption{Dependence of the traffic flow distribution on the distance covered. The distance shown is between two cities, not exits.}
\label{fig:mileage}
\end{center}
\end{figure}

\begin{figure}
\begin{center}
\includegraphics[width=0.8\textwidth]{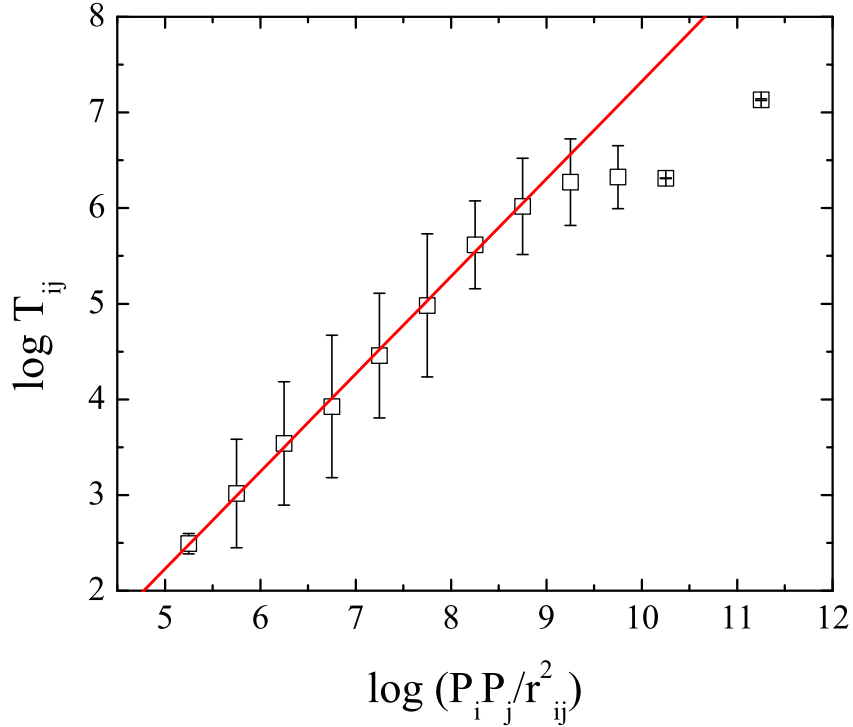}
\caption{The gravity model for the Korean highway system. The parameters, population $P$ and distance $r$, lead to the metaphor of physical gravity as described in Newton's law of gravity. The slope of a guide line is 1.02.}
\label{fig:gravity}
\end{center}
\end{figure}

Fig. \ref{fig:population} represents the correlation between the population, the area and the total (the above panels), in and out traffic flows (the bottom panels). The population and the traffic flows indicate positive correlations; the exponent is 0.95 (total), 0.93 (in), and 0.95 (out), on a log-log scale. However, the traffic flows of selected cities are not proportionate to the area. It is not only because the total traffic flow of a given city is more related to the population than the area, but also because the area and the population have no correlation for Korean cities. 

When a driver travels from one city to another, the route might be flexible depending on a driver's personality, traffic condition, and so on. In addition, a city might have a lot of exits, but they are regarded as one exit. Therefore, it is not easy to measure the actual distance covered by a car. We define the distance between two cities as the distance covered. The average of the distribution of distance $d$ between two cities is 238.9km, and the standard deviation is 128.5km. Fig. \ref{fig:mileage} shows the correlation between the distribution of traffic flows $P(T)$ and distance $d$. The correlation can be classified into two parts. The slope of the left guideline of Fig. \ref{fig:mileage}, represents the first part, which is 0.00189, and that of the right one is 0.01555. Considering the average of distance, 238.9$\pm$128.5km, the left part covers the average and its error bar region. In this part, the traffic flow depends very slightly on distance compared to the right part. However, this result might be regarded as a finite-size effect by considering the fact that Korea is not a big country.

We investigate traffic flows for selected 30 cities in terms of both population and distance. Fig. \ref{fig:gravity} represents the histogram of traffic flows between city $i$ and $j$, $T_{ij}$, as a function of the two populations $P_i$ and $P_j$ of city $i$ and $j$ and $r_{ij}$ the distance between $i$ and $j$. It is found that they form a gravity model \cite{erlander1990B,haynes1984B,zipf1947},

\begin{equation}
T_{ij}\sim \frac{P_i P_j}{r_{ij}^2}.
\label{eq:gravity1}
\end{equation}

Some spots, which are the first left and four right spots, are removed from linear fitting in Fig. \ref{fig:gravity}. It is hard to analyze for these spots since they do not contain enough number of data. That may be due to the finite-size effect as mentioned earlier. Furthermore, there could have been another possibility. These spots almost correspond to the traffic flows among Seoul and Suwon, Yongin, and Ansan. These cities are very close. Seoul is the metropolis and the largest city of Korea. Suwon and Yongin have large industrial factories such as Samsung Electronics Company, whose capitalization occupies over 20\% of the Korean stock market \cite{jung2006}. Additionally, there is a famous industrial complex in Ansan. Korea is highly centralized in the metropolitan area than other countries. This phenomenon is a serious problem for Korea. Practically, these cities consist of the Greater Seoul. This centralization is also observed in Fig. \ref{fig:map}. Most Korean cities are located near Seoul and Busan, the second largest city. 

\section{Network analysis}
In this section, we establish the highway network using the whole 238 exits data. In other words, a node of the network in this section corresponds to an \textit{exit}. The degree distribution $P(k)$, a commonly used as a parameter to measure the network properties, is not meaningful for fully connected network such as the highway system which is analyzed here. Moreover, clustering coefficient $c$ and average path length $l$ are not used to analyze the network in this work. However, the weight of a link $w_ij$ and the strength $s_i$, defined as

\begin{equation}
s_i=\sum^N_{j=1}w_{ij},
\end{equation}

\noindent can be a significant measure of the network properties.

\begin{table}
\begin{center}
\begin{tabular}{|c|c||c|c|}
\hline
Exit&Ratio&Exit&Ratio\\
\hline
Seoul&0.05099&Hwawon&0.01600\\
West Seoul&0.03980&North Daegu&0.01458\\
East Seoul&0.03128&Busan&0.01455\\
Gunja&0.02793&Jangyu&0.01304\\
North Busan&0.02529&North Suwon&0.01279\\
West Daegu&0.02388&Cheonan&0.01164\\
Suwon&0.02344&Osan&0.01130\\
West Busan&0.02070&Daejeon&0.01128\\
Daedong&0.01989&East Daegu&0.01121\\
East Suwon&0.01864&East Gimhae&0.01083\\
\hline
\end{tabular}
\caption{\label{table:strength} List of twenty exits which have largest strength, $s_i$. The \textit{ratio} represents $s_i/\sum_j s_j$. Seoul, Busan, and Daegu have four exits in this table, respectively. Three are included in Suwon, and Jangyu and East Gimhae are exits of Busan's satellite city.}
\end{center}
\end{table}

\begin{figure}
\begin{center}
\includegraphics[width=0.45\textwidth]{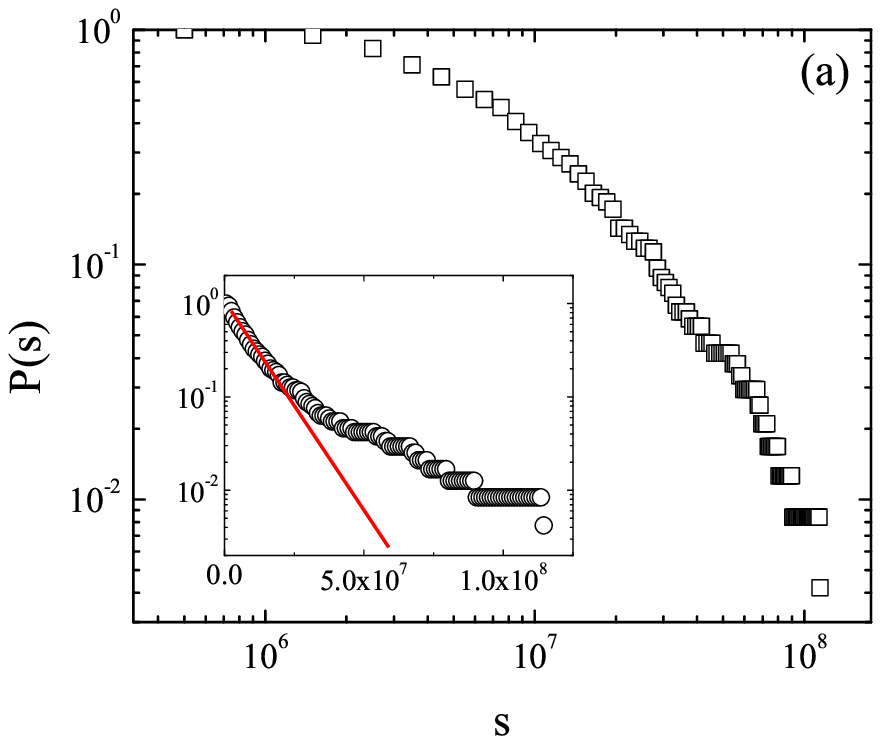}
\includegraphics[width=0.45\textwidth]{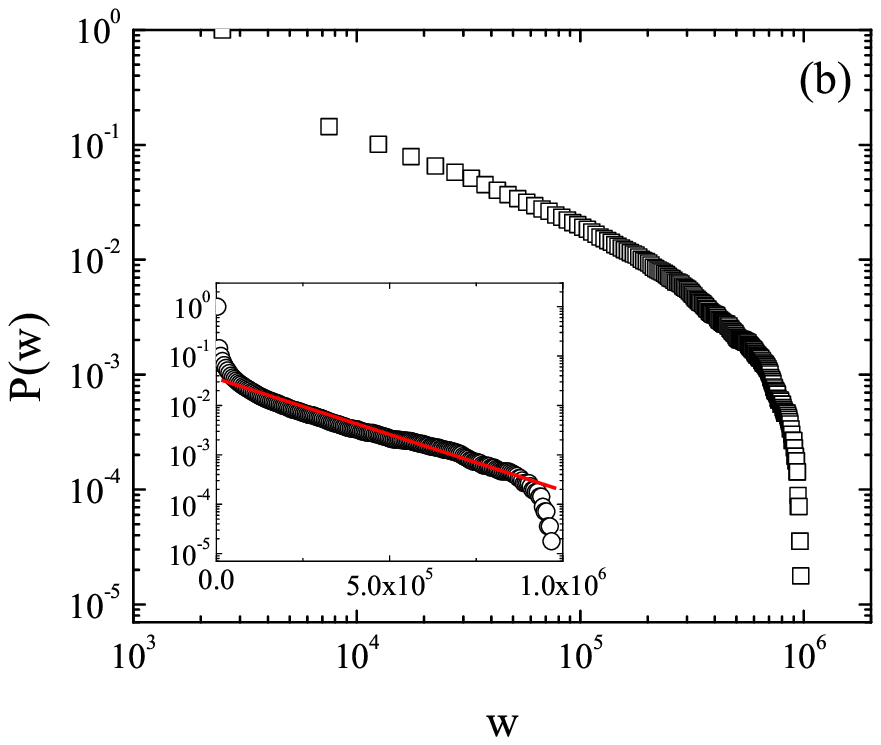}
\caption{Cumulative probability distribution of (a) strength and (b) weight for the whole 238 exits for the whole cars on log-log scales (insets represent semi-log scales). The distribution of strength shows heavy tail, but not power-law behavior. The lines serve as a guide to the eyes.}
\label{fig:strength}
\end{center}
\end{figure}

Fig. \ref{fig:strength} represents the cumulative probability distributions of (a) strength and (b) weight for the whole 238 exits data on log-log and semi-log (insets) scales. The strength distribution shows a \textit{heavy tail} on a semi-log plot, but not the power-law behaviors. In addition, the weight distribution is not a power-law. The spots consisting of the heavy tail in Fig. \ref{fig:strength}(a) correspond to big cities such as Seoul, Busan, and so on (Table \ref{table:strength}). Gunja is an entrance of eastbound near Seoul, Daedong is a part of Busan, and Hwawon is that of Daegu. It is typical that the strength is proportional to the population because the strength is the sum of traffic flows related into the population. 

However, the distributions of degree and strength for the Worldwide Airport Network \cite{barrat2004} show power-law behaviors. Basically as well as practically, road system is different from airport system. Usually, there are more limitations for constructing and expanding routes on land than in the air, physically. This property allows more powerful hubs and heavy links for air transportation network than ground transportation. Therefore, road network is usually regarded as a rather \textit{uniform} and \textit{homogeneous} network, and airport system as an \textit{inhomogeneous} one \cite{jeong2003}.  Airport network as well as air transportation network establishes a scale-free network, an inhomogeneous network model with power-law distribution \cite{barrat2004}. Furthermore, the \textit{public} ground transportation network has the power-law behavior for its weight distribution \cite{montis2005} because it has preferential attachment and powerful hubs, which are important characteristics of the scale-free network model \cite{barabasi2002B}, rather than the private transportation. Transportation companies make more preferential attachments, powerful hubs, and a lot of traffic flows between hubs than personnel. 

We go back to the Korean highway system, which contains both \textit{public} and \textit{private} transportation information. The heavy tail of Fig. \ref{fig:strength}(a) shows that through the network has some powerful hubs, it is not sufficient to show power-law behaviors. It is natural to imagine that a lot of traffic flows connect big cities like the public ground transportation. But road system cannot be an inhomogeneous network because of the limitation described earlier. Therefore, heavy tail is observed, but not sufficient to show the power-law behavior. Moreover, the width of the road cannot be infinite, and the weight distribution does not have the heavy tail.

\section{Conclusions}
We investigated the Korean highway system. A gravity model as the metaphor of physical gravity is found in the system. The original Newton's law of gravity is $F\sim M_1 M_2/r^2$, but the gravity model for the Korean highway system can be derived as $T\sim P_1 P_2/r^2$, where $P$ denotes population. Air and public ground transportation network, which have preferential attachment properties, have scale-free behaviors such as power-law distribution. However, road network is basically a rather uniform and homogeneous one such as random network. Public ground transportation network which formed on road network is based on the revenue for companies. Therefore, public transportation network has inhomogeneous property. The Korean highway network including public and private transportation information shows heavy tail, but not power-law distribution. Private transportation also tends to connect larger cities, but road network limits the system to be inhomogeneous.

\end{document}